\documentclass[aps,pra,a4paper,twocolumn,noshowpacs,superscriptaddress,floatfix]{revtex4}
\usepackage{amsfonts}
\usepackage{amsmath}
\usepackage{amssymb}
\usepackage{array}
\usepackage{graphicx}
\usepackage{epstopdf}
\usepackage{multirow}
\usepackage{booktabs}
\usepackage{makecell}
\usepackage{setspace}

\newcommand{\PreserveBackslash}[1]{\let\temp=\\#1\let\\=\temp}
\newcolumntype{C}[1]{>{\PreserveBackslash\centering}p{#1}}
\newcolumntype{R}[1]{>{\PreserveBackslash\raggedleft}p{#1}}
\newcolumntype{L}[1]{>{\PreserveBackslash\raggedright}p{#1}}

\begin{document}

\title{An application of the Hylleraas-B-splines basis set: High accuracy calculations of the static dipole polarizabilities of helium}

\author{San-Jiang Yang}
\affiliation{Department of Physics, Wuhan University, Wuhan 430072, China}
\affiliation{State Key Laboratory of Magnetic Resonance and Atomic and Molecular Physics, Wuhan Institute of Physics and Mathematics,
Chinese Academy of Sciences, Wuhan 430071, People’s Republic of China}

\author{Xue-Song Mei}
\affiliation{Department of Physics, Wuhan University, Wuhan 430072, China}

\author{Ting-Yun Shi \footnote{electronic mail: tyshi@wipm.ac.cn}}
\affiliation{State Key Laboratory of Magnetic Resonance and Atomic and Molecular Physics, Wuhan Institute of Physics and Mathematics,
Chinese Academy of Sciences, Wuhan 430071, People’s Republic of China}
\affiliation{Center for Cold Atom Physics, Chinese Academy of Sciences, Wuhan 430071, People’s Republic of China}

\author{Hao-Xue Qiao \footnote{electronic mail: qhx@whu.edu.cn}}
\affiliation{Department of Physics, Wuhan University, Wuhan 430072, China}

\begin{abstract}
The Hylleraas-B-splines basis set is introduced in this paper, which can be used to obtain the eigenvalues and eigenstates of helium-like system's Hamiltonian. Comparing with traditional B-splines basis, the rate of convergence of our results has been significantly improved. Through combine this method and pseudo-states sum over scheme, we obtained the high precision values of static dipole porlarizabilities of the $1{}^1S-5{}^1S$, $2{}^3S-6{}^3S$ states of helium in length and velocity gauges respectively, and the results get good agreements. The final extrapolate results of porlarizabilities in different quantum states arrived eight significant digits at least, which fully illustrates the advantage and convenience of this method in the problems involving continuous states.
\end{abstract}
\maketitle

\section{INTRODUCTION}
The accurate calculations and experimental measurements of physical effects for helium led to a sustained attention because of its many applications. For instance, compared with hydrogen, the 2P state of helium has a longer life-time which can be used for the more precise determination of the fine-structure constant \cite{schwartz1964fine}\cite{pachucki2010fine}\cite{feng2015laser}\cite{pei2015precision}. On the other hand, high accuracy results of oscillator strengths and polarizabilitiers of helium could give out the response of helium to the external field, and provide a new route to test the non-relativity quantum electrodynamics(NRQED) \cite{PhysRevA.88.052515}\cite{henson2015precision}\cite{zhang2016tune}.

Helium as a three-body system, its eigenvalue problem does not have an analytic expression. To get the high precision approximations, the Rayleigh-Ritz variational method is usually adopted. The basic idea of variational calculations is to construct one convenient and efficient basis set which could contain some optimized parameters and should be nearly-completeness when the number of basis vector is large enough. In 1929, Hylleraas got a successful attempt in the calculation of the energy levels of helium due to he accounted for the correlation between the two electrons in helium and constructed Hylleraas-type basis set \cite{hylleraas1929neue}. The effectiveness of this basis set aroused interests of numerous researchers, as an outstanding review, Drake's article showed us the stories of Hylleraas-type development \cite{drake2006springer}. On the way of solving the atomic calculation problems in different aspects, the researchers developed a number of excellent basis sets. For instance, the doubled Hylleraas basis set can effectively enhance the convergence rate of energy levels of the Rydberg states of helium \cite{drake1988high}; the Hylleraas-Gaussian basis set can be used to accurately calculate the lower energy levels of helium in strong magnetic field \cite{PhysRevA.77.043414}; through explicitly including powers of logarithmic and half-integral terms in Hylleraas basis set, the variational non-relativistic energy level of the ground state of Helium can even achieved up to 36 digits \cite{doi:10.1142/S0218301306004648}.

The variational method we have mentioned above is usually not considered for the continuum states, but when one wants to study the polarizabilities of an atom, or to calculate the energy levels shift when it comes to some higher-order QED correction items, the continuum states should be carefully considered. For this situation, the idea of pseudo-states is concise and effective. In the article \cite{doi:10.1139/p00-010}, Drake and Goldman gave an exquisite treatment of the continuum states. However, we adopt the following ideas. In our calculation scheme, first we put the atom in one limited-size, non-penetrable box, so the Hamiltonian spectrum is discrete. In this case, we can get the values of the physical effects which we are interested (contains the contribution of the positive energy state to it). Then we let the size of the box gradually increased, what we could expect is that when the box is large enough, the result we get should be close to the exact value of the nature sufficiently.

In order to utilize the pseudo-states concept, we should find one proper basis set which is generated in a finite-size space. In the various basis sets which are usually adopted in atomic and molecular calculation, the B-splines basis set has the good performance \cite{0034-4885-64-12-205}\cite{FroeseFischer20111315}. As every B-spline in this basis set is high local and generated in one finite-size box, so to combine this type basis set with the pseudo-states concept is natural. By utilizing the concept of the pseudo-states, the dipole polarizabilities of helium have been studied in details by using the B-splines basis set \cite{chen1995dipole}\cite{PhysRevA.92.012515}. By analysing the data in this articles, we find that due to the B-spines basis set doesn't reflect the behaviour of near region of the two electrons, the calculation of porlarizabilitis requires rather large number of partial waves. The accuracy of the results using the B-splines basis is less accurate than the one using Hylleraas-type basis. In order to improve the flexibility of B-splines in the calculation of helium, by coupling the correlated term $r_{12}=\left| \vec r_1-\vec r_2 \right|$ to the traditional B-splines basis set, we constitute a more effective basis set, called Hylleraas-B-splines (H-B-splines, see expression (\ref{H-B-splines})). The H-B-splines basis set inherits the virtues of the traditional B-splines, and it does not suffer from the numerical linear-dependence problem. Because of the correlated term directly appears in the H-B-splines basis, the rate of convergence of partial waves improved greatly.

In this article, we calculated the energy levels and static dipole porlarizabilitis of the $1{}^1S - 5{}^1S$ and $2{}^3S - 6{}^3S$ of helium using H-B-splines basis. In the results of energy levels, for the ground state which is an obvious bottleneck to be precisely calculated by traditional B-splines basis, our result convergent to 11 significant digits rapidly. For the calculation of static dipole porlarizabilitis, although the energy levels of the initial states are not as precise as one get from Hylleraas basis, as the H-B-splines basis has the more satisfactory capacity to simulate the entire intermediate energy levels, the results of static dipole porlarizabilitis of helium in our calculation arrived 8 significant digits at least in length gauge. More details numerical results are arranged in the section \uppercase\expandafter{\romannumeral3}. Atomic unites is used throughout in this paper.

\section{THEORETICAL METHOD}
\subsection{HYLLERAAS-B-SPLINES}
Helium is a three-body system which consists of one nucleus of charge $2e$, mass M and two electrons of charge $-e$, mass $m$. Its reduced non-relativistic Hamiltonian is
\begin{equation}
     H=\sum_{i} \left( -\frac{1}{2\mu} \vec \nabla_i^2 -\frac{1}{r_i} \right) -\frac{1}{M}
     \vec \nabla_1 \cdot \vec \nabla_2 + \frac{1}{r_{12}}\ ,
\end{equation}
where $\mu=mM/(m+M)$ is the electron reduced mass, $r_i$ is the distance between the i-th electron and nucleus, $r_{12}$ is the distance of two electrons. For simplicity, we set the $M$ to be infinite in our calculations.

To use the Rayleigh-Rize variational method, by coupling the $r_{12}$ and B-splines basis, we construct the H-B-splines basis set as below
\begin{equation}
\label{H-B-splines}
\begin{aligned}
\{ \Phi _{ijcl_1l_2} =& B_{i,k} \left( r_1 \right) B_{j,k} \left( r_2 \right) r_{12}^c \\& \mathbf{\Lambda}_{l_1l_2}      ^{LM} \left( \hat{r}_1 ,\hat{r}_2 \right) \pm exchange \} \qquad \left( 0,r_{max} \right).
\end{aligned}
\end{equation}
Here $\mathbf{\Lambda}_{l_1l_2}^{LM}$ is the vector coupled product of angular momenta $l_1$, $l_2$ for the two electrons
\begin{equation}
\begin{aligned}
   \mathbf{\Lambda}_{l_1l_2}^{LM} \left( \hat{r}_1 ,\hat{r}_2 \right) =
    \sum_{m_1,m_2} \langle l_1l_2m_1m_2 \mid LM \rangle \\
    Y_{l_1m_1}(\hat{r}_1) Y_{l_2m_2}(\hat{r}_2) \ ,
\end{aligned}
\end{equation}
and the rest parts are consisting of B-splines functions $B_{i,k}(r)$ \cite{boor1978a} and correlated term $r_{12}$. The $i, k$ denote the serial number and the order of $B_{i,k}(r)$ respectively, the shape of the function $B_{i,k}(r)$ is depend on the non-decreasing knots sequence $\{t_k \}$ and its order. The $r_{max}$ is the size of the box, which leads the B-splines are generated in a finite space.

The B-splines satisfy the recursive relation
\begin{equation}
    B_{i,k}(r) = \frac{r-t_i}{t_{i+k-1}-t_i} B_{i,k-1}(r) +
    \frac{t_{i+k}-r}{t_{i+k}-t_{i+1}} B_{i+1,k-1}(r) ,
\end{equation}
together with the definition of B-splines of order $k=1$
\begin{equation} \label{eq}
 \left\{ \begin{aligned}
           B_{i,k}(r)&=1 \qquad t_i \leq r < t_{i+1},\\
           B_{i,k}(r)&=0 \qquad otherwise .
        \end{aligned}
 \right.
\end{equation}
In the calculation of helium, the exponential sequence always be used
\begin{equation}
  \lambda_i =r_{min} + (r_{max} - r_{min}) \frac{e^{\gamma \left( \frac{i-1}{n-1} \right)  }-1}{e^{\gamma}-1}\ .
\end{equation}
It's convenient to set $\gamma =\alpha \times r_{max} $. To satisfy the boundary conditions of the wave functions of helium, the knots sequence should be arranged as below:
\begin{equation}
\left\{ \begin{aligned}
           t_i&=0 \qquad \qquad \qquad \qquad \quad   i=1,2,\ldots ,k-1,\\
           t_i&=r_{max}  \frac{e^{\gamma \left( \frac{i-k}{N-k+1} \right)  }-1}{e^{\gamma}-1} \quad i=k,k+1,\ldots ,N,\\
           t_i&=r_{max} \qquad \qquad \qquad \quad \ i=N+1,\ldots ,N+k-1.
        \end{aligned}
 \right.
\end{equation}

If the $r_{max}$, knots sequence, order $k$ and total number $N$ of B-splines have been determined, the basis set of H-B-splines is completely constructed.

In the choice of parameters $i,j,c,l_1,l_2$, we restrict c to less than two, therefore the polynomials of $\cos \theta $ from correlated terms will not directly present in basis set, which means that the correlated terms are not in conflict with the freedom of the choice of $l_1, l_2$, so the parameters should be arranged as bellow,
\begin{equation}
\begin{aligned}
           i&=1,2,\ldots ,j ,  \\
           j&=1,2,\ldots ,N  , \\
           c&=0,1 , \\
           l_1&=0,1,\ldots ,l_{max} , \\
           l_2&=0,1,\ldots ,l_{max} .
        \end{aligned}
\end{equation}
As the wave function must be anti-symmetric about the index of two electrons, the term which leads the normal of $\Phi_{ijcl_1l_2}$ to be zero must be eliminated.

\subsection{POLARIZABILITY}

The averaged static $2^l$-pole polarizability for an atom is defined as
\begin{equation}
\overline{\alpha}_l =\sum_{n \neq 0} \frac{\overline{f_{n0}^{(l)}}}{(E_n - E_0)^2} \ ,
\end{equation}
where $\overline{f_{n0}^{(l)}}$ is the averaged $2^l$-pole oscillator strength defined by
\begin{equation}
\begin{aligned}
\overline{f_{n0}^{(l)}} =&\frac{8\pi}{(2l+1)^2(2L+1)} (E_n - E_0) \\& \left| \left\langle
\psi_0 \left\| \sum_i r_i^l Y_{lm}(\hat{r}_i) \right\| \psi_n \right\rangle \right|^2 .
\end{aligned}
\end{equation}
the $i$ denotes the serial number of electrons in an atom, $E_n$ and $\psi_n$ denote the intermediate energy levels and corresponding states, $E_0$ and $\psi_0$ denote the eigenvalue and the associated eigenstate which we interested.

For dipole polarizability, using the commutation relation
\begin{equation}
\begin{aligned}
(E_n-E_0)&\langle 0 |\vec{r}_i | n\rangle = \\& \langle 0 |[\vec{r_i},H]| n\rangle =-i \langle 0 |\vec{p}_i| n \rangle,
\end{aligned}
\end{equation}
one can get another form (velocity gauge) of the averaged dipole oscillator strength as below
\begin{equation}
\begin{aligned}
\overline{f_{n0}^{(1)}}=\frac{8\pi}{(2l+1)^2(2L+1)(E_n - E_0)} \\ \left| \left\langle
\psi_0 \left\| \sqrt{\frac{3}{4\pi}}\sum_i \vec p_1 \cdot \vec{e}_m \right\| \psi_n \right\rangle \right|^2 ,
\end{aligned}
\end{equation}
where $\vec{e}_m$ denotes the three unit vectors with $m=0,\pm1$ defined by
\begin{equation}
\vec{e}_{\pm 1}=\mp \frac{1}{\sqrt{2}} (  \vec{e}_x \pm \vec{e}_y          ),
\vec{e}_0=\vec{e}_z
\end{equation}

In different gauges, the polarizabilities must be the same value for the same state. Therefore, we calculate in two gauges respectively, and through the relative difference of two gauges to ensure the correctness of our results, which defined by
\begin{equation}
\eta=2 \cdot \frac{|a-b|}{a+b},
\end{equation}
where $a$, $b$ are the results from two gauges respectively.

\section{RESULTS AND DISCUSSIONS}
Figure 1 presents the convergence test of the polarizabilities of $2{}^3S$ of helium as the size of the box $r_{max}$ increased. In this calculations, we set the number of the partial waves, $l_{max}$, equal to one, the $\gamma=0.038$.

\begin{figure}
\centering
\includegraphics[scale=.32]{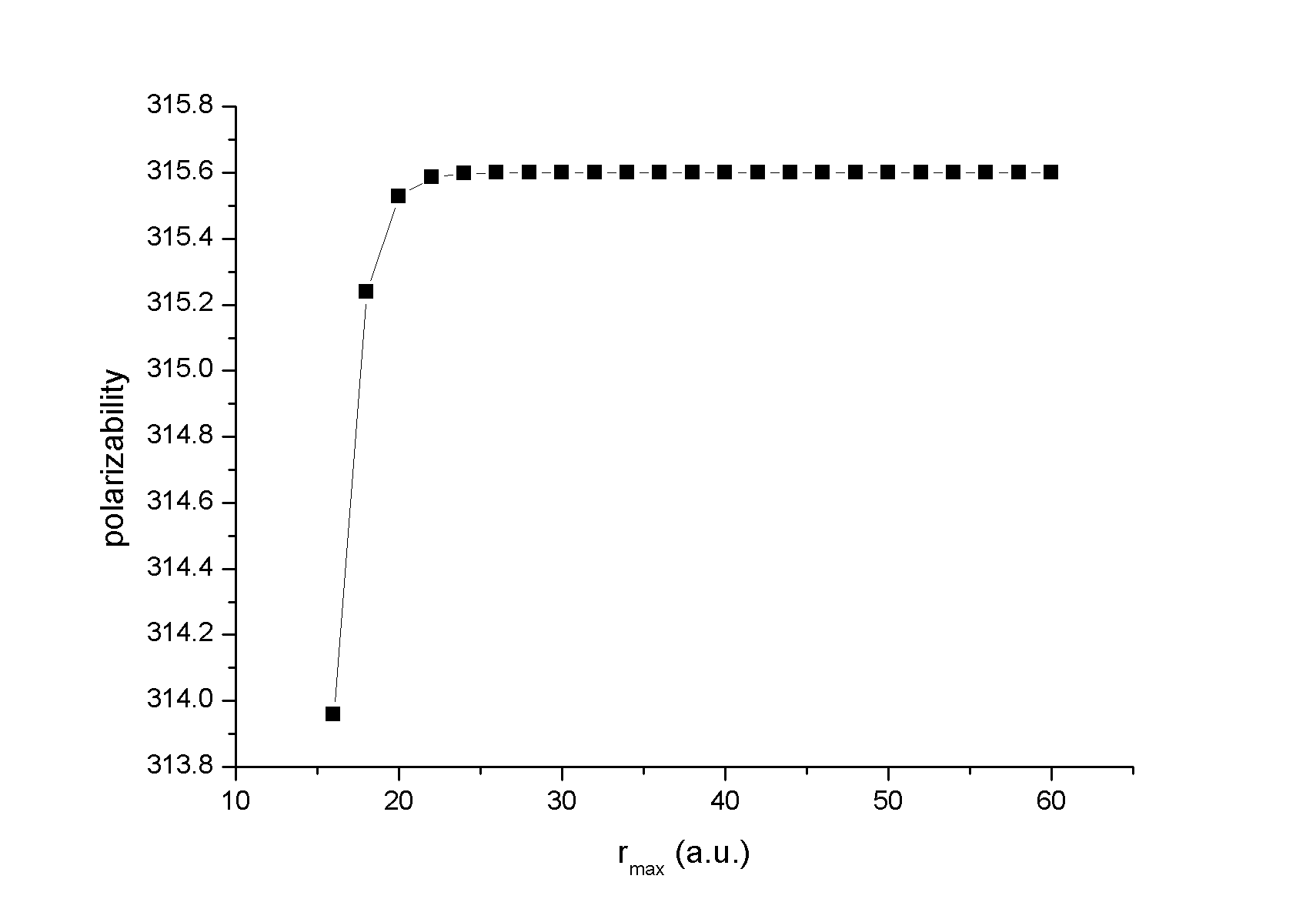}
\caption{Convergence of the polarizabilities of $2{}^3S$ of helium as the size of the box $r_{max}$ increasing, the $l_{max}=1$, $N=50$. The black points represent the results which we calculated. Units are in a.u.}
\end{figure}

From the Figure 1, we could see the variation tendency of polarizablities, that is, when we increase the $r_{max}$ up to about 30 a.u., the curve is quite flat. In our calculation, we set $r_{max}=200$ a.u., since the interested states $5^1S$ and $6^3S$ have comparable high principle quantum number.

\subsection{ENERGY LEVELS}
Table \uppercase\expandafter{\romannumeral1} shows the convergence of the energy of state $1^1S$, with different partial wave $l$ along with an increasing $N$. Our extrapolated value is $E=-2.9037243770(1)$, 4 significant digits better than $E=-2.903724268$ getting from B-splines basis \cite{rao1997method} using $l_{max}=16$ . As we have mentioned the improve of the rate of convergence of partial waves compare with using the B-splines basis, our results reach up to the same significant digits in B-splines basis using $l_{max}=16$ just need using $l_{max}=1$, which shows the effectiveness of the addition of the correlated term. The extrapolated results of $2^3S$ state and the other states are listed in Table \uppercase\expandafter{\romannumeral2}. A comparation with other results are also presented.

As can be seen in Table \uppercase\expandafter{\romannumeral2}, comparing with B-splines basis (third column), the accuracy of our results is significantly improved. The results are excellently agree with the other high accuracy results which getting from the exponential correlated Hylleraas basis (fourth column) and doubled Hylleraas basis (fifth column) respectively. Not only for the ground state but for the states of high principal quantum number, our results remain high accuracy, which implies that using H-B-splines basis could be another effective way to calculate the energy levels of the high principal quantum number.

\subsection{STATIC DIPOLE POLARIZABILITIES}
In Table \uppercase\expandafter{\romannumeral3} and Table \uppercase\expandafter{\romannumeral4}, we give the convergence studies of the polarizabilities of $2^3S$ in length and velocity gauges respectively as the number of $N$ and $l_{max}$ increasing. The final convergent value in length gauge is 315.6314723(2) which has 3 significant digits more than 315.6315(2) of article \cite{PhysRevA.92.012515}, moreover, our calculation need $l_{max}=4$ instead of $l_{max}=10$. In Table \uppercase\expandafter{\romannumeral5}, we list the relative difference $\eta$ of the data of Table \uppercase\expandafter{\romannumeral3} and Table \uppercase\expandafter{\romannumeral4} in different $N$ and $l_{max}$. Obviously, the discrepancy of polarizabilities in two gauges is smaller and smaller with $N$ and $l_{max}$ increasing. When $N=50$ and $l_{max}=4$, the relative difference $\eta=6 \times 10^{-11}$. We give the extrapolated results of the $1{}^1S-5{}^1S$, $2{}^3S-6{}^3S$ states in two gauges in Table \uppercase\expandafter{\romannumeral6}, and the relative difference $\eta$ is $9 \times 10^{-10}$ at most in these calculations, which fully ensured the correctness of our results on convergence. Table \uppercase\expandafter{\romannumeral6} also contains the corresponding results from other works.

In Table \uppercase\expandafter{\romannumeral6}, similar to the calculation of energy levels, the numerical accuracy of static dipole polarizabilities is improved obviously compared with B-splines basis (fourth column). In second and third column, we list the results of polarizabilities in two gauges respectively. Obviously, the results of two sets are quite self consistency, and the results arrived 8 significant digits at least. Comparing with the high accuracy results from doubled Hylleraas basis (fifth column), one can find that, for the lower energy states, $1^1S$, $2^1S$, $3^1S$, $2^3S$ and $3^3S$, good agreements are archived in two basis, however, for the rest states, the results of two basis appear some differences. Since the two forms mutually verified in our calculations, we think our results are more reliable. As the significant digits of our results are rarely decreasing with the raising principle quantum number, this could offer an efficient way to calculate the static dipole polarizabilities of Rydberg atoms with high accuracy.

\section{SUMMARY}
In this article, we constructed the H-B-splines basis by coupling and the correlated term $r_{12}=\left| \vec r_1-\vec r_2 \right|$ to the traditional B-splines basis. This basis has good capacity to describe the near region behaviour of the two electrons. As we have shown in the article, the H-B-splines basis not only improves the computational accuracy and efficiency about excited states of helium, but also makes the calculation of the ground state no longer to be the bottleneck of the traditional B-splines basis. On the other hand, this basis is particularly suitable for the problem involving continual states as every B-splines in it is high local.

In our calculations, our results of energy levels and static dipole polarizabilities of helium arrived 11 and 8 significant digits at least, which are not easy to achieve by using traditional the B-spline basis. Especially for polarizabilities, the accuracy of our results is comparable with the high accuracy results of the doubled Hylleraas basis, which indicates that H-B-splines basis provides another effective way for the mutual verification of different basis sets. It's worth to mention that, since the accuracy of our results do not deteriorate with the principal quantum number increasing, H-B-splines basis leads another way to calculate the energy levels and polarizabilities of Rydberg states of helium-like atoms.

Furthermore, in many computational occasions of helium-like atoms, the concept of intermediate states appears frequently, for example, the quadrupole and octupole polarizabilities and the computation of Bethe Logarithms or the black body radiation shifts \cite{zhou2016black}. Through the calculations in this article, the validity of H-B-splines basis about these situations is predictable.

\section{ACKNOWLEDGMENTS}
The authors thank Yong-Bo Tang and Wan-Ping Zhou for meaningful discussions. This work was supported by the National Natural Science Foundation of China (No.11674253) and (No.91536120).

\makegapedcells
\setcellgapes{4pt}
\begin{table*}[!htbp]
\caption{Convergence of the energy of the $1{}^1S$ states of helium as the number of B-splines N and partial waves $l_{max}$ increased. The numbers in parentheses of the extrapolated values give the
computational uncertainties. Units are in a.u.}
\begin{tabular}{C{1cm}L{3.5cm}L{3.5cm}L{3.5cm}L{3.5cm}L{3.5cm}}
\hline
\hline
N&\qquad $l_{max}=1$&\qquad $l_{max}=2$&\qquad $l_{max}=3$&\qquad $l_{max}=4$\\
\hline
20&-2.9036&-2.90371&-2.90371&-2.90371\\
25&-2.903723&-2.903723&-2.903723&-2.9037240\\
30&-2.9037241&-2.90372430&-2.90372434&-2.90372435\\
35&-2.90372425&-2.90372436&-2.903724371&-2.903724373\\
40&-2.903724265&-2.903724375&-2.903724375&-2.9037243764\\
45&-2.903724267&-2.9037243766&-2.9037243767&-2.90372437690\\
50&-2.9037242683&-2.90372437687&-2.90372437696&-2.903724376999\\
Extrap.&&&&-2.9037243770(1)\\
\hline
\hline
\end{tabular}
\end{table*}

\makegapedcells
\setcellgapes{4pt}
\begin{table*}[!htbp]
\caption{Comparison of the energies for the five lowest singlet and the five lowest triplet states of helium. The numbers in parentheses of the extrapolated values give the
computational uncertainties. Units are in a.u.}
\begin{tabular}{C{1.5cm}C{3.5cm}C{3.5cm}C{3.5cm}C{4cm}C{3.5cm}}
\hline
\hline
state&this work&Ref.\cite{chen1994accurate}&Ref.\cite{cann1992oscillator}&Ref.\cite{drake1994variational}\\
\hline
$1{}^1S$&-2.9037243770(1)&-2.9035774&------&-2.9037243770341195\\
$2{}^1S$&-2.14597404605(1)&-2.1459649&-2.1459740292&-2.145974046054419(6)\\
$3{}^1S$&-2.061271989737(4)&-2.0612681&-2.0612719720&-2.061271989740911(5)\\
$4{}^1S$&-2.033586717027(5)&-2.0335850&-2.0335866995&-2.03358671703072(1)\\
$5{}^1S$&-2.02117685157(1)&-2.021175&-2.0211768309&-2.021176851574363(5)\\

$2{}^3S$&-2.17522937823(1)&-2.1752288&-2.175229378176&-2.17522937823679130\\
$3{}^3S$&-2.068689067469(5)&-2.0686888&-2.068689067283&-2.06868906747245719\\
$4{}^3S$&-2.036512083095(5)&-2.0365120&-2.036512082933&-2.03651208309823630(2)\\
$5{}^3S$&-2.022618872299(5)&-2.0226188&-2.022618871382&-2.02261887230231227(1)\\
$6{}^3S$&-2.015377452989(4)&------&-2.015377452422&-2.01537745299286219(3)\\
\hline
\hline
\end{tabular}
\end{table*}

\makegapedcells
\setcellgapes{4pt}
\begin{table*}[!htbp]
\caption{Convergence of the static dipole polarizabilities in length gauge for the $2{}^3S$ states of helium as the number of B-splines N and partial waves $l_{max}$ increased. Units are in a.u.}
\begin{tabular}{C{1cm}L{3.5cm}L{3.5cm}L{3.5cm}L{3.5cm}L{3.5cm}}
\hline
\hline
N&\qquad $l_{max}=1$&\qquad $l_{max}=2$&\qquad $l_{max}=3$&\qquad $l_{max}=4$\\
\hline
20&315.5&315.6&315.6311&315.637\\
25&315.601&315.6311&315.6315&315.6316\\
30&315.6004&315.63148&315.631475&315.63148\\
35&315.600336&315.63147&315.6314722&315.631473\\
40&315.6003313&315.631465&315.6314723&315.6314726\\
45&315.6003316&315.6314644&315.6314724&315.6314724\\
50&315.600331943&315.631464209&315.631472397&315.631472384\\
Extrap.&&&&315.6314723(2)\\
\hline
\hline
\end{tabular}
\end{table*}

\makegapedcells
\setcellgapes{4pt}
\begin{table*}[!htbp]
\caption{Convergence of the static dipole polarizabilities in velocity gauge for the $2{}^3S$ states of helium as the number of B-splines N and partial waves $l_{max}$ increased. Units are in a.u.}
\begin{tabular}{C{1cm}L{3.5cm}L{3.5cm}L{3.5cm}L{3.5cm}L{3.5cm}}
\hline
\hline
N&\qquad $l_{max}=1$&\qquad $l_{max}=2$&\qquad $l_{max}=3$&\qquad $l_{max}=4$\\
\hline
20&315.3&315.5&315.61&315.62\\
25&315.5550&315.62&315.6311&315.6313\\
30&315.55591&315.63140&315.63143&315.63146\\
35&315.55590&315.631455&315.63146&315.6314723\\
40&315.555913&315.631452&315.631471&315.6314724\\
45&315.555916&315.6314508&315.6314722&315.631472364\\
50&315.55591865&315.631450648&315.631472353&315.63147236368\\
Extrap.&&&&315.6314723(1)\\
\hline
\hline
\end{tabular}
\end{table*}

\makegapedcells
\setcellgapes{4pt}
\begin{table*}[!htbp]
\caption{Relative differences $\eta$ of the static dipole polarizabilities of $2^3S$ of helium between the two specifications at different $N$ and $l_{max}$. Units are in a.u.}
\begin{tabular}{C{1cm}L{3.5cm}L{3.5cm}L{3.5cm}L{3.5cm}L{3.5cm}}
\hline
\hline
N&\qquad $l_{max}=1$&\qquad $l_{max}=2$&\qquad $l_{max}=3$&\qquad $l_{max}=4$\\
\hline
20&0.0006&0.0001&0.00006&0.00003\\
25&0.0001&0.000004&0.000001&0.0000009\\
30&0.0001&0.0000002&0.0000001&0.00000006\\
35&0.0001&0.00000005&0.00000001&0.000000004\\
40&0.0001&0.00000004&0.000000002&0.0000000006\\
45&0.0001&0.00000004&0.0000000006&0.0000000001\\
50&0.0001&0.00000004&0.0000000001&0.00000000006\\
\hline
\hline
\end{tabular}
\end{table*}

\makegapedcells
\setcellgapes{4pt}
\begin{table*}[!htbp]
\caption{Comparison of the static dipole polarizabilities for the five lowest singlet and the five lowest triplet states of helium. The numbers in the parentheses are the
computational uncertainties. Units are in a.u.}\label{tab:tablenotes}
\begin{tabular}{C{2cm}C{3.5cm}C{3.5cm}C{3.5cm}C{3.5cm}}
\hline
\hline
state&this work(length gauge)&this work(velocity gauge)&Ref.\cite{chen1995dispersion}&Ref.\cite{yan2000polarizabilities}\\
\hline
$1{}^1S$&1.383192174(2)&1.383192174(1)&1.38328&1.38319217440(5)\footnote{this data form Ref.\cite{yan1996variational}}\\
$2{}^1S$&800.316233(2)&800.31623(1)&800.306&800.31633(7)\\
$3{}^1S$&16887.1856(1)&16887.1856(2)&&16887.17(1)\\
$4{}^1S$&135851.581(1)&135851.581(1)&&135851.430(1)\\
$5{}^1S$&669586.06(1)&669586.06(2)&&669585.8982(2)\\

$2{}^3S$&315.6314723(2)&315.6314723(1)&315.630&315.63147(1)\\
$3{}^3S$&7937.58592(2)&7937.58592(1)&&7937.58(1)\\
$4{}^3S$&68650.2089(2)&68650.208(1)&&68650.061(2)\\
$5{}^3S$&351796.229(2)&351796.22(1)&&351796.060(2)\\
$6{}^3S$&1314954.97(1)&1314954.9(1)&&1314954.806(3)\\
\hline
\hline
\end{tabular}
\end{table*}

\bibliography{H-B-splines}

\end{document}